\documentclass[times,10pt,twocolumn]{article} 
\usepackage{latex8}
\usepackage{times}
\usepackage[T1]{fontenc}
\usepackage{graphics}
\makeatletter
\newcommand{\LyX}{L\kern-.1667em\lower.25em\hbox{Y}\kern-.125emX\@}
\newenvironment{lyxcode}
  {\begin{list}{}{
    \setlength{\rightmargin}{\leftmargin}
    \raggedright
    \setlength{\itemsep}{0pt}
    \setlength{\parsep}{0pt}
    \ttfamily}%
   \item[]}
  {\end{list}}
\usepackage[T1]{fontenc}
\usepackage[latin1]{inputenc}
\usepackage{graphics}



\begin{document}
\title{Replica Selection in the Globus Data Grid}

\author{Sudharshan Vazhkudai,$^1$ Steven Tuecke,$^2$ and
Ian Foster$^2$\\
\\
 $^1$ \textit {Department of Computer and Information Science}\\
 \textit {The University of Mississippi}\\
 \textit {chucha@john.cs.olemiss.edu}\\
\\
 $^2$ \textit {Mathematics and Computer Sciences Division}\\
 \textit {Argonne National Laboratory}\\
 \textit{\{tuecke, foster\}@mcs.anl.gov}\normalsize }

\maketitle
\thispagestyle{empty}
\begin{abstract}
The Globus Data Grid architecture provides a scalable infrastructure for the
management of storage resources and data that are distributed across Grid environments.
These services are designed to support a variety of scientific applications,
ranging from high-energy physics to computational genomics, that require access
to large amounts of data (terabytes or even petabytes) with varied quality of
service requirements. By layering on a set of core services, such as data transport,
security, and replica cataloging, one can construct various higher-level services.
In this paper, we discuss the design and implementation of a high-level replica
selection service that uses information regarding replica location and user
preferences to guide selection from among storage replica alternatives. We first
present a basic replica selection service design, then show how dynamic information
collected using Globus information service capabilities concerning storage system
properties can help improve and optimize the selection process. We demonstrate
the use of Condor's ClassAds resource description and matchmaking mechanism
as an efficient tool for representing and matching storage resource capabilities
and policies against application requirements. 
\end{abstract}

\section{Introduction}

An increasing number of scientific applications ranging from high-energy physics
to computational genomics require access to large amounts of data---currently
terabytes and soon petabytes---with varied quality of service requirements.
This diverse demand has contributed to the proliferation of storage system capabilities,
thus making storage devices an integral part of the Grid environment and thereby
constituting the Data Grid. The Globus \cite{Globus00} Data Grid architecture
is an effort to standardize access to the multitude of storage systems spread
across the Grid environment. It attempts to abstract these diverse elements
of the data grid by providing a set of core services that can be used to construct
a variety of higher-level services \cite{DataGrid}.

In this paper, we first describe the basic Globus Data Grid architecture, explaining
briefly the various components. We then describe the process of identifying
characteristics of interest about a storage replica resource, and we illustrate
the mechanism with which these characteristics can be published. Next, we explain
the architecture of replica selection, a higher-level service, built by using core
services provided by the Globus data grid. We conclude by presenting
the use of Condor's classified advertisements (ClassAds) and matchmaking mechanisms~\cite{Livny98}
as an elegant matching and ranking tool in a storage context.

\section{Architecture of Data Grid}

The Globus data grid is organized into two layers, namely,
core services and higher-level services that are built using the core services.
The working hypothesis is that this hierarchical organization will make it possible
to reuse services and code across a variety of applications and tools,
so that, for example, both application-specific replica management solutions
and sophisticated storage management systems such as the Storage Request Broker (SRB) \cite{SRB} can share common low-level mechanisms.

\begin{figure*}
{\par\centering {\includegraphics{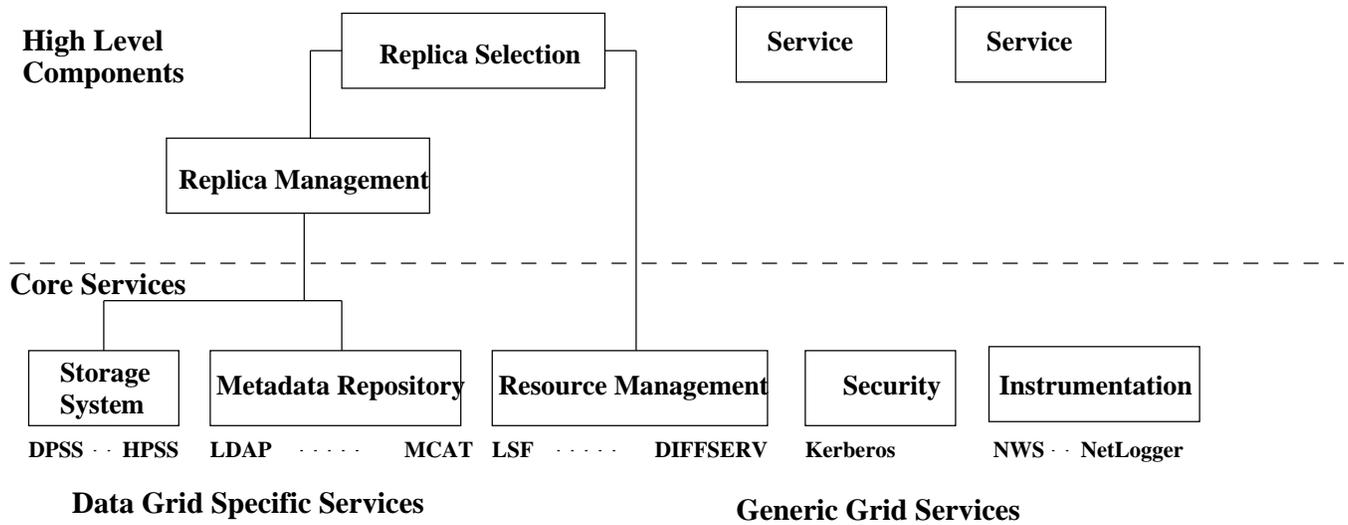}} \par}

\caption{Data Grid Architecture \label{DataGridArch}\cite{DataGrid}.}
\end{figure*}
\subsection{Core Services}

Core Data Grid services (Figure \ref{DataGridArch}) seek to abstract the multitude
of storage systems that exist in Grid environments, so that higher-level services---and
applications---can access those storage systems uniformly. These core services
include the following:

\begin{itemize}
\item \textbf{Storage Systems and Data Access} - Storage system and data access services
provide basic mechanisms for accessing and managing the data located in storage
systems. These mechanisms provide abstractions for uniformly creating, deleting,
accessing and modifying file instances across storage systems, regardless of
their physical location. They target secondary and archival storage systems
such as Unix file systems, High Performance Storage System (HPSS) \cite{HPSS},
and Unitree and can also be used to access more sophisticated systems such
as SRB.
\item \textbf{Metadata Access} - The metadata access service provides mechanisms to
access and manage information about the data stored in storage systems. Various
kinds of metadata are provided by this service:

\begin{enumerate}
\item \textbf{Application Metadata} describes the contents of the file, the circumstances
under which the data was collected, and various details used by the application. 
\item \textbf{Replica Metadata} describes the mapping between file instances and particular
replica locations. 
\item \textbf{System Configuration Metadata} describes the capabilities of storage
systems, thus providing information on the fabric of the Data Grid. 
\end{enumerate}
The metadata service further provides a uniform method with which the various
metadata can be published and accessed \cite{DataGrid}.

\end{itemize}
These services themselves build on basic security and information services provided
by the Globus Toolkit \cite{Globus00}.

\subsection{Higher-Level Services}

The core Data Grid services can be used to construct a variety of higher-level
services (Figure \ref{DataGridArch}). For example:

\begin{itemize}
\item \textbf{Replica Management} - Replica management is the process of creating
or deleting replicas at a storage site. Most often, these replicas are exact
copies of the original files, created only to harness certain performance benefits.
A replica manager typically maintains a replica catalog containing replica site
addresses and the file instances. 
\item \textbf{Replica Selection} - Replica selection is the process of choosing a
replica from among those spread across the Grid, based on some characteristics
specified by the application. One common selection criteria would be access
speed \cite{DataGrid}. 
\end{itemize}

\section{Storage System Functionality and Replication}

Intelligent replica selection requires information about the capabilities and
performance characteristics of a storage system \cite{DataGrid}. We address
this information discovery problem by leveraging machinery provided by the Globus
Metacomputing Directory Service (MDS) \cite{mds97}, an information collection,
publication, and access service for Grid resources.

The Globus MDS uses the Lightweight Directory Access Protocol (LDAP) \cite{Howes97}
as its access protocol and LDAP object classes as its data representation, but
it adopts innovative approaches to the problems of resource registration and discovery.
Information about an individual resource or set of resources is collected and
maintained by a Grid Resource Information Service (GRIS) daemon, which responds
to LDAP requests with dynamically generated information and can be configured
to register with one or more Grid Index Information Services (GIISs). Users
will typically direct broad queries to GIIS to discover resources and then
drill down with direct queries to GRIS to get up-to-date, detailed information
about individual resources.

\begin{figure*}
{\par\centering {\includegraphics{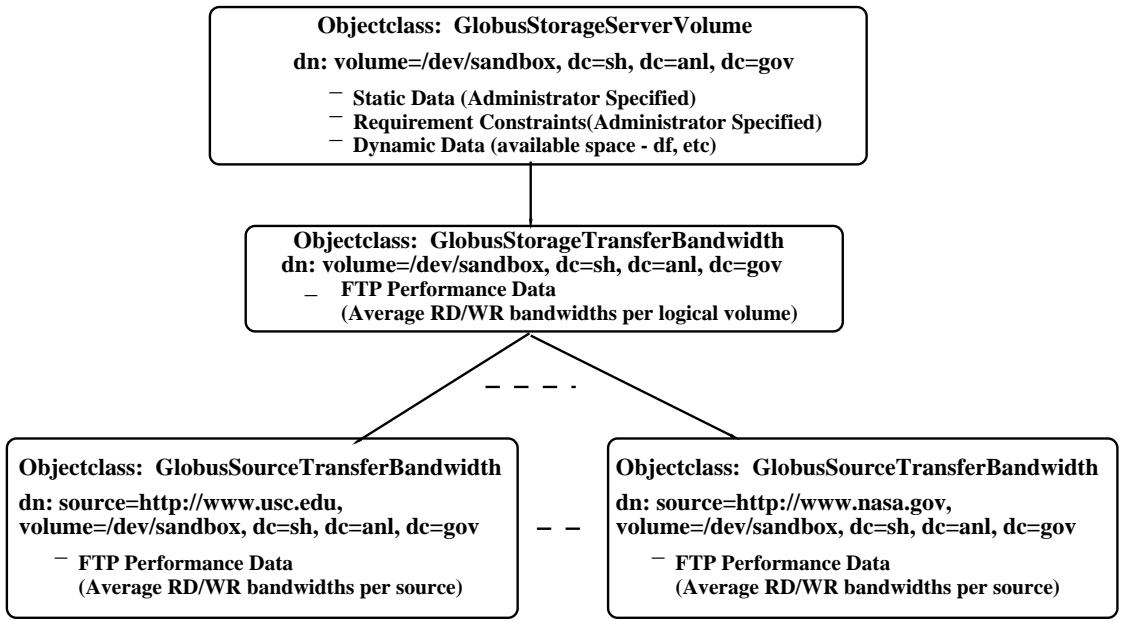}} \par}

\caption{The directory information tree structure used in our storage system GRIS\label{DIT}}
\end{figure*}

Information in LDAP is organized in a tree structure referred to as a Directory
Information Tree (DIT), with nodes in a DIT tree corresponding to the LDAP structured
data types called object classes. Figure~\ref{DIT} depicts the DIT hierarchy
of object classes that we have defined to describe storage systems in a Data
Grid environment. We describe these object classes in the following section.

\subsection{Storage GRIS}

Each storage resource in the Globus Data Grid incorporates a Grid Resource Information Server, configured to collect and publish system configuration metadata
describing that storage system. This information typically includes attributes
such as storage capacity and seek times, and descriptions of site-specific policies
governing storage system usage. Figure \ref{SysConf} shows the object class
definition that we have developed for this data. Attributes are labeled as
either ``MUST CONTAIN'' or ``MAY CONTAIN,'' indicating whether or not they
are mandatory.

\begin{figure}
\centering 
\begin{tabular}{|l|}
\hline 
Grid::Storage::ServerVolume \\
 OBJECT CLASS ::=\{\\
 SUBCLASS OF Grid::Physical Resource\\
 RDN = gss(Grid::Storage::ServerVolume)\\
 CHILD OF \{\\
 \ \ \ Grid::organizationalUnit\\
 \ \ \ Grid::organization\\
 \ \ \ Grid::Top\\
 \}\\
 MUST CONTAIN \{\\
 \ \ \ totalSpace::cisfloat::singular,\\
 \ \ \ availableSpace::cis::singular,\\
 \ \ \ mountPoint::cisfloat::singular,\\
 \ \ \ diskTransferRate::cisfloat::singular,\\
 \ \ \ drdTime::cisfloat::singular,\\
 \ \ \ dwrTime::cisfloat::singular,\\
 \ \ \ {}\}\\
 MAY CONTAIN \{\\
 \ \ \ requirements::cis::singular,\\
 \ \ \ filesystem::cis::multiple,\\
 \}\\
 \} \\
\hline 
\end{tabular}
\par{}

\caption{System Configuration Metadata specification using LDAP object classes.\label{SysConf}}
\end{figure}

This object class contains both dynamic and static attributes: attributes such
as \texttt{totalSpace}, \texttt{availableSpace}, and \texttt{mountPoint} are
dynamic, varying with various frequencies; attributes such as \texttt{diskTransferRate},
\texttt{drdTime}, \texttt{dwrTime}, and \texttt{requirements} are more static.

The requirements attribute is particularly interesting: it allows an administrator
to specify the conditions under which the specified storage replica can be used,
based, for example, on device utilization. For example, the requirements attribute
could be a Boolean expression of maximum allowable storage and transfer bandwidth
and can be specified using the ClassAd \cite{Livny98} mechanism. We discuss
ClassAds in detail in a later section.

The data shown in Figure \ref{SysConf} is gathered in a variety of ways. We
base our general approach on a generic LDAP-based GRIS server developed within
the Globus project. The OpenLDAP server has a feature by which shell scripts
(``shell-backends'') can be executed at the back-end in response to search
queries; we use such scripts to gather dynamic attributes such as \texttt{availableSpace},
\texttt{mountPoint}, and \texttt{totalSpace}. Static attributes such as site
usage policies and seek times can be specified by the administrator in a configuration
file. Data collected in this manner is published in a suitable format (for example,
LDIF \cite{Howes97}).

\subsection{Data Access Service}

We are also interested in the speed of a storage system, or, rather, in the time
that the storage system can be expected to take to deliver a replica. One approach
to determining this information is to construct a performance model of the relevant
components (e.g., see \cite{Chowdry}). We favor an alternative approach in
which historical information concerning data transfer rates is used as a predictor
of future transfer times. In brief, storage systems are configured to provide
information on their own behavior and performance. Attributes such as maximum
achievable read and write transfer bandwidths across networks can help an application
choose one replica over another. Such data can be obtained by the storage replica
by monitoring their own performance. This feature can be extended further, to
obtain statistical information based on the performance data, such as average
transfer bandwidths and their standard deviations, that can help predict the
behavior of a particular replica. Further, we expect that there will be significant
reuse of storage servers by clients, thereby justifying performance information
on a per source basis, which provides a client useful information on end-to-end
transfer performance between the server and the client. For example, a simple
heuristic of combining past observed performance with current load of server
might give a client a reasonably good choice of server.

Figures \ref{TransBand} and \ref{SrcTransBand}, along with Figure \ref{DIT},
depict the object classes used to record performance data associated with a
replica location. The object class in Figure \ref{TransBand} specifies a summary
of transfer bandwidth performance for all replica transfers, and the object
class in Figure \ref{SrcTransBand} specifies the performance details from the
replica location to particular source sites. We gather this performance data
by using instrumentation incorporated in the GridFTP server \cite{Globus00}\cite
{DataIntense}.

\section{Classified Advertisements}

\label{classads}

Classified Advertisements (ClassAds) \cite{Livny98} are an elegant matching
mechanism used in the Condor high-throughput computing system to map resource
capabilities against job requirements. ClassAds allow resources and jobs to
advertise their capabilities and requirements as attribute-value pairs that
can be matched. Two ClassAds match if the logical expressions contained in the
``requirements'' attribute in both of them is satisfied. This requirements
attribute itself can be represented in terms of other attributes. Further, ClassAds
provide a mechanism for ranking the matches based on some attribute value, or
value computed from multiple attributes. Up until now, ClassAds have been used
extensively in Condor for job placements. In this section, we present the use
of ClassAds in a storage context.

The various storage attributes described in the preceding section can be represented
as attribute-value pairs in a ClassAd. For example, the following is a simple
ClassAd describing the capabilities of a storage resource:

\begin{lyxcode}
hostname~=~``hugo.mcs.anl.gov'';

volume~=~``/dev/sandbox'';

availableSpace~=~50G;

MaxRDBandwidth~=~75K/Sec;

requirement~=~other.reqdSpace~<~10G~

\&\&~other.reqdRDBandwidth~<~75K/Sec;~
\end{lyxcode}
This ClassAd describes a volume of a storage resource by specifying its attributes.
The ClassAd also specifies a usage policy enforced by the resource, whereby
only applications requiring storage less than 10 GB and transfer bandwidths
less than 75 KB/sec are granted access. When two ClassAds are being matched,
a MatchClassAd is created that contains both ClassAds. Each ClassAd can refer
to the other ClassAd by using the ``other'' keyword.

A ClassAds representation of storage capabilities provides an efficient environment
for matching, querying, and ranking requests. We will discuss the specifics of
a request ClassAd and the matching process once we describe the replica selection
mechanism.
\begin{figure}
\centering 
\begin{tabular}{|l|}
\hline 
Grid::Storage::TransferBandwidth \\
 OBJECT CLASS ::=\{\\
 SUBCLASS OF Grid::StorageServerVolume\\
 RDN = gss(Grid::TransferBandwidth)\\
 CHILD OF \{\\
 \ \ \ Grid::StorageServerVolume\\
 \ \ \ Grid::organizationalUnit\\
 \ \ \ Grid::organization\\
 \ \ \ Grid::Top\\
 \}\\
 MUST CONTAIN \{\\
 \ \ \ MaxRDBandwidth::cisfloat::singular,\\
 \ \ \ MinRDBandwidth::cisfloat::singular,\\
 \ \ \ AvgRDBandwidth::cisfloat::singular,\\
 \ \ \ MaxWRBandwidth::cisfloat::singular,\\
 \ \ \ MinWRBandwidth::cisfloat::singular,\\
 \ \ \ AvgWRBandwidth::cisfloat::singular,\\
 \ \ \ ..........................................................\\
 \}\\
 \} \\
\hline 
\end{tabular}
\par{}

\caption{Performance data specification describing entire site characteristics, using
LDAP object classes. \label{TransBand}}
\end{figure}

\section{Replica Selection}

As noted above, a Data Grid provides a convenient environment for a community
of researchers, interested in particular data sets, to maintain replicas of
the data sets at their respective sites. Such an environment would provide both faster
access and better performance characteristics. Replica selection is a high-level
service provided by the Data Grid based on the core services. The replica selection
process allows an application to choose a replica, from among those in a replica
catalog, based on its performance and data access features.

An application that requires access to replicated data begins by querying an
application specific metadata repository, specifying the characteristics of
the desired data. The metadata repository maintains associations between representative
characteristics and logical files, thus enabling the application to identify
logical files based on application requirements. Once the logical file has been
identified, the application uses the replica catalog to locate all replica locations
containing physical file instances of this logical file, from which it can choose
a suitable instance for retrieval.

The entity that identifies the suitable instance of a replicated file based
on application requirements is referred to as a broker. In effect, the responsibility
of the broker is to map application requirements against storage resource capabilities.
In the following section, we discuss the storage broker in detail.

\begin{figure}
\centering 
\begin{tabular}{|l|}
\hline 
Grid::Storage::SourceTransferBandwidth \\
 OBJECT CLASS ::=\{\\
 SUBCLASS OF Grid::TransferBandwidth\\
 RDN = gss(Grid::SourceTransferBandwidth)\\
 CHILD OF \{\\
 \ \ \ Grid::StorageTransferBandwidth\\
 \ \ \ Grid::StorageServerVolume\\
 \ \ \ Grid::organizationalUnit\\
 \ \ \ Grid::organization\\
 \ \ \ Grid::Top\\
 \}\\
 MUST CONTAIN \{\\
 \ \ \ lastWRBandwidth::cisfloat::singular,\\
 \ \ \ lastWRurl::cis::singular,\\
 \ \ \ lastRDBandwith::cisfloat::singular,\\
 \ \ \ lastRDurl::cis::singular,\\
 \}\\
 \} \\
\hline 
\end{tabular}
\par{}

\caption{Performance data specification describing per source characteristics using
LDAP object classes. \label{SrcTransBand}}
\end{figure}
\subsection{The Storage Broker}

In this section, we first delve into the details of the storage broker architecture.
We then discuss the overall architecture, a decentralized selection process, and
a few issues involved in its realization.

\subsubsection{Decentralized Selection Process}

Traditionally, resource brokers have adopted a centralized approach to resource
management, wherein a single node is responsible for decision making. An example
of such an environment is the Condor \cite{Condor00} high-throughput computing
platform, wherein a central manager is responsible for matching resources against
jobs. Obvious disadvantages to this approach are scalability and a single point
of failure. Of course, Condor has an efficient recovery mechanism to address
failure and has been proven to scale to thousands of resources and users.

But there is a more fundamental problem with this centralized approach when
applied to Grids. In these highly distributed environments, there are numerous
user communities and shared resources, each with distinct security requirements.
No single resource broker is likely to be trusted by all of these communities
and resources with the necessary information to make decisions. At the extreme,
each user may need his or her own resource broker, because only that user has
the authorization to gather all of the information necessary to make brokering
decisions.

For this reason, we have designed a decentralized storage brokering strategy
wherein every client that requires access to a replica performs the selection
process rather than a central manager performing matches against clients and
replicas.

\subsubsection{Architecture}

Figure \ref{Broker} presents a snapshot of the Grid environment where storage
resources are scattered and each client requiring access to a replica initiates
a decentralized replica selection mechanism. As can be seen, there is no central
point of control, and decision making is delegated to each and every client.

An application requiring access to a file presents its requirements as a ClassAd
to the broker. These requirements might be as simple as a Boolean expression
indicating the storage and transfer bandwidth required or may indicate more
complex constraints on storage system type, state, and policy. The broker then
performs the following sequence of actions:

\begin{itemize}
\item \textbf{Search Phase}

\begin{enumerate}
\item The broker attempts to find a suitable replica matching the application's requirements.
To achieve this, it queries the replica catalog, which contains addresses
of all replicas for each logical file. 
\item The next step is to query each replica location. As we have seen before, this
involves using LDAP searches to query GRIS servers associated with storage systems. 
\item In response to the LDAP search, each storage system returns its capabilities
and usage policies in the LDAP Information Format. 
\item The broker collects the capabilities of all replica resources and proceeds to
the matching phase. 
\end{enumerate}
\item \textbf{Match Phase}

\begin{enumerate}
\item The replica capability data is converted into ClassAd format
in preparation for invoking the Condor matchmaking mechanisms. We thus obtain
a list of classified advertisements, representing the various replica sites. 
\item The broker then performs a match of the application's requirement ClassAd against
the list of replica capability ClassAds, obtaining a set of replica locations
that satisfy the criterion. 
\item The ClassAd ranking feature can be used to prioritize successful matches based
on some attribute, specified by the application. 
\end{enumerate}
\item \textbf{Access Phase}

\begin{enumerate}
\item Once a suitable replica has been identified, the file is accessed using a high-speed
file transfer protocol, for example the GridFTP tools provided within the Globus
Toolkit. 
\begin{figure*}
{\par\centering {\includegraphics{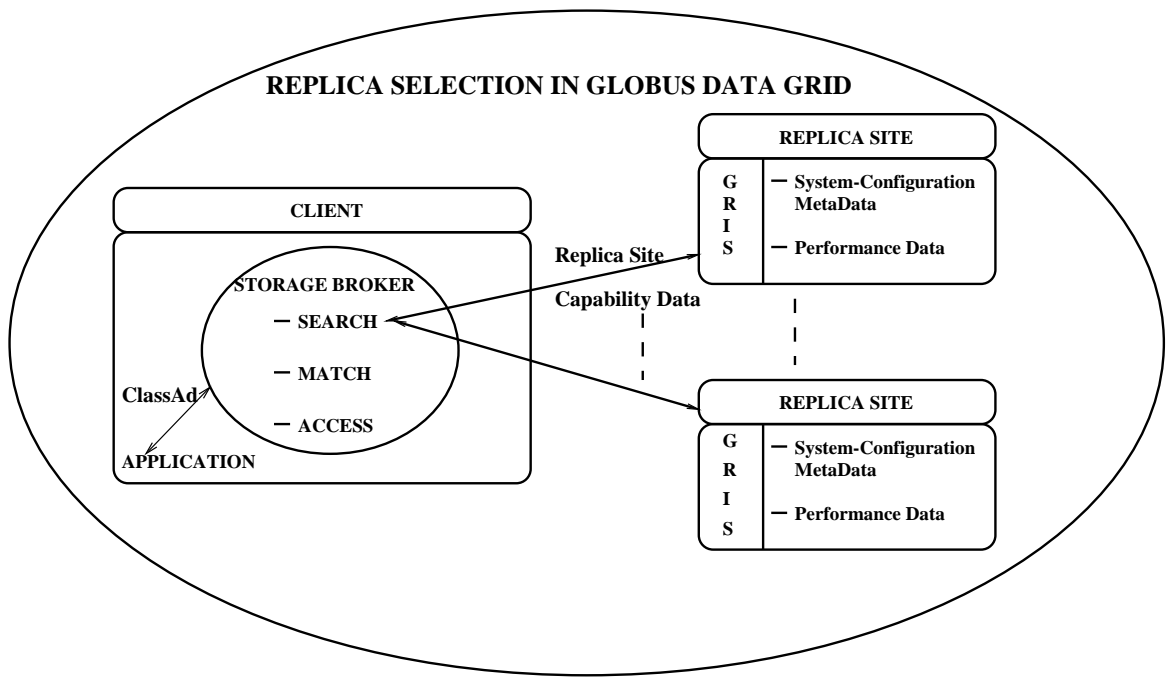}} \par}

\caption{Storage Broker Architecture \label{Broker}.}
\end{figure*}

\end{enumerate}
\end{itemize}

\subsection{An Example Application Request}

In Section \ref{classads}, we described a storage ClassAd that indicates its
willingness to accept application requests that require space less than 10 GB
and require a transfer bandwidth less than 75 KB/sec. In this section, we look
at a user request ClassAd and how it is matched against the storage ClassAd.

An application might advertise its request to the broker as follows:

\begin{lyxcode}
hostname~=~``comet.xyz.com'';

reqdSpace~=~5G;

reqdRDBandwidth~=~50K/Sec;

rank~=~other.availableSpace;

requirement~=~other.availableSpace~>
~~~5G~ \&\&~other.MaxRDBandwidth~>
~~~50K/Sec;~
\end{lyxcode}
The application indicates its preference for a storage resource that has more
than 5 GB available space and a maximum transfer bandwidth greater that 50 KB/sec.
The broker, upon receiving such a request, attempts to contact all replica locations
identified by the replica catalog, using LDAP search queries to request the
attributes of interest: in this case, \texttt{availableSpace} and \texttt{MaxRDBandwidth}.
The broker thus uses the application ClassAd to build specialized LDAP search
queries.

The results obtained via these LDAP queries are converted from LDAP Interchange
Format (LDIF) into ClassAds and standard Condor mechanisms are invoked to match
the application ClassAd with each ClassAd in the list of storage system ClassAds.
(Note that the storage system ClassAds comprise their site specific usage policy,
as mentioned in the sample storage ClassAd in Section \ref{classads}.) Any
matched ClassAd can be further ranked by querying their rank attribute. In our
case, we rank the replica servers based on their available space, thus obtaining
the ``best'' match \cite{Livny98}.

\section{Results}

We summarize our initial results based on our experience in building
this prototype.

We have succeeded in building a replica selection broker and have demonstrated
that this high-level Data Grid service can be built using the services
provided by the Globus Data Grid Toolkit, such as: Storage GRIS, GridFTP and Replica Catalog.

We have further demonstrated the process of: Identifying characteristics of interest on a storage resource, based on Storage GRIS and GridFTP protocols;
Classifying them as belonging to appropriate object classes according to LDAP;
Gathering these attributes, via such mechanisms as shell backend scripts, tuning
FTP servers and configuration files;
Publishing the features in a suitable, efficiently queriable format using LDAP
protocol.

We have explored the use of Condor ClassAds as a mechanism for expressing
storage resource capabilities in the Grid environment. Although LDAP provides
an equivalent method of publishing characteristics in attribute-value pairs,
ClassAds provides a richer matching and ranking environment. Further, in our prototype implementation, we have demonstrated that the process of converting
data, represented in LDAP format, into ClassAds is not cumbersome and is worth
the effort. We have, in fact, developed primitive libraries to achieve the conversion
of this attribute set.

\section{Conclusions and Future Directions}

Our current prototype implementation is primarily a proof of concept of the
Globus Data Grid services. As the next logical step in its development, we
need to demonstrate its applicability in a real application, thereby corroborating
its usefulness. Further improvements can be made with regards to the information
published in the storage GRIS. Finally, the statistical information published
by the storage resource can be fed to an information service, such as the Network
Weather Service \cite{NWS}, to perform predictive analysis of the behavior
of storage resources.

\section*{Acknowledgments}

This work was supported in part by the Mathematical, Information, and Computational Sciences Division subprogram of the Office of Advanced Scientific Computing
Research, U.S. Department of Energy, under Contract W-31-109-Eng-38; by the National Science Foundation; and by the NASA Information Power Grid program.

\bibliographystyle{plain}
\bibliography{literature}

\end{document}